# Atomic-scale visualization of *d*-wave altermagnetism

Daran Fu,[1,2†] Liu Yang,[3†] Yi Shen,[3] Kebin Xiao,[1,2] Yuyang Wang,[1,2] Wei Jiang,[3,4,5] Zhiwei Wang,[3,4,5*] Yugui Yao,[3,4,5] Qi-Kun Xue,[1,2,6,7,8*] and Wei Li[1,2,8*]

[1]*State Key Laboratory of Low-Dimensional Quantum Physics, Department of Physics, Tsinghua University, Beijing 100084, China*

[2]*Frontier Science Center for Quantum Information, Beijing 100084, China*

[3]*Centre for Quantum Physics, Key Laboratory of Advanced Optoelectronic Quantum Architecture and Measurement (MOE), School of Physics, Beijing Institute of Technology, Beijing 100081, China*

[4]*Beijing Key Lab of Nanophotonics and Ultrafine Optoelectronic Systems, Beijing Institute of Technology, Beijing 100081, China*

[5]*International Center for Quantum Materials，Beijing Institute of Technology, Zhuhai, 519000, China*

[6]*Beijing Academy of Quantum Information Sciences, Beijing 100193, China*

[7]*Southern University of Science and Technology, Shenzhen 518055, China*

[8]*Hefei National Laboratory, Hefei 230088, China*

*To whom correspondence should be addressed: zhiweiwang@bit.edu.cn; qkxue@mail.tsinghua.edu.cn; weili83@tsinghua.edu.cn

Altermagnetism is a newly identified magnetic phase, distinct from conventional ferromagnetism and antiferromagnetism. It exhibits no net magnetization while breaks time-reversal symmetry. Although its momentum-space signatures are established, direct real-space visualization of its defining rotational-symmetry breaking remains missing. Here, using scanning tunnelling microscopy, we provide atomic-scale real-space evidence for altermagnetism in $CsV_2Se_2O$. Utilizing intrinsic spin defects as probes, we directly visualize the hallmark symmetry breaking through unidirectional electronic patterns and elliptical charging rings, both tied to the alternating spin texture. Moreover, adjacent spin-defect lines exhibit opposite spins and long-range antiferromagnetic coupling, hinting at a novel spin order. Our work moves the field from momentum-space probes to direct real-space visualization, opening a path to explore how this unconventional magnetic order couples to other quantum states.

Altermagnetism has emerged as a third elementary category of magnetic order, completing the established dichotomy of ferromagnetism and antiferromagnetism[1-3]. Its defining characteristic is the coexistence of a vanishing net magnetization, akin to antiferromagnets, with a broken time-reversal symmetry, a hallmark of ferromagnets[2,4-7]. This unique blend arises from specific crystal symmetries where the arrangement of non-magnetic atoms imposes an environment that forces a crystallographic rotation, rather than a simple translation, to connect the spin-up and spin-down sublattices (Fig. 1a, b)[1,8]. This underlying symmetry breaking is predicted to produce alternating spin polarizations in both real and momentum space, leading to novel phenomena such as non-zero Berry curvature, anomalous Hall effects and momentum-dependent spin splitting, which have spurred significant research interest for both fundamental physics and potential applications[2-22].

The formal classification of this phase relies on the non-relativistic spin group theory, which treats transformations in spin space and real space independently[3,23-27]. In this framework, altermagnetism is described by a combined operation $[R_i||R_j]$, coupling a rotation in spin space to a distinct rotation in real space[3]. This differs fundamentally from conventional ferromagnetism (a single spin lattice) or antiferromagnetism (spin sublattices linked by translation or inversion). The consequential reduction in point group symmetry from the crystal lattice to each spin sublattice, for example, from four-fold ($C_4$) of the crystal square lattice to two-fold ($C_2$) of the spin sublattice, is the root cause of the anisotropic, alternating spin-polarized bands observed in reciprocal space[1,2,9,28-34]. Despite these well-established momentum-space signatures and theoretical predictions, the direct and visual proof, the real-space observation of its broken rotational symmetry in the electronic structure at the atomic scale, has been a critical missing piece.

$CsV_2Se_2O$, a predicted *d*-wave altermagnet[2], provides an exemplary platform to address this gap. Its $V_2O$ plane forms an anti-$CuO_2$ lattice[28,35] (Fig. 1a, b), making it a magnetic counterpart[1,36,37] to high-temperature superconducting cuprates and an ideal system to explore *d*-wave order originating from crystal symmetry without strong electron correlations. While previous studies on related compounds have confirmed key momentum-space features like alternating spin splitting and $C_2$-symmetric Fermi surface (FS, Fig. 1c)[28], the real-space electronic landscape is uncharted. In this work, we employ scanning tunneling microscopy (STM) to investigate $CsV_2Se_2O$ at the atomic scale. We directly image the broken rotational symmetry through unidirectional static charge orders and electronic charging rings encircling spin defects. Quasiparticle interference (QPI) measurements independently reveal scattering wavevectors that confirm the alternating spin-polarized band structure. Furthermore, we uncover a spin-density-wave (SDW) gap and an associated charge density wave (CDW). Collectively, these combined observations offer the long-sought, direct real-space verification of *d*-wave altermagnetism.

While a spin-polarized tip directly measures spin polarization[38,39], it is not the only route to access the spin degree of freedom. Our approach instead relies on an intrinsic probe: the spin defect. By symmetry, these defects are inherently tied to a specific spin sublattice, therefore serving as a probe to directly visualize the reduced symmetry. In $AV_2Se_2O$ ($A$ = K, Rb, Cs) materials, each $V_2O$ layer is sandwiched by two Se layers that share the same square lattice as the spin-up and spin-down sublattices (Fig. 1a, b), rendering direct real-space detection of the spin texture exceptionally challenging. Using a conventional tip, our results provide the first firm and abundant evidence for real-space symmetry lowering in altermagnetism, resolving charge-channel characteristics that ultimately originate from the underlying spin texture. Our method thus moves beyond inferring spin symmetry; it visualizes, at the atomic scale, how the spin and charge degrees of freedom actively interact within the altermagnetic state, a fundamental aspect that has remained experimentally elusive.

**Magnetic structure and SDW gap**

$CsV_2Se_2O$ crystallizes in a layered tetragonal structure (Fig. 1a and Extended Data Fig. 1). Its spin configuration in the $V_2O$ plane leads to a vanishing net magnetization (Extended Data Fig. 2), similar to an antiferromagnet. Se atoms reside above and below the center of each $V_2O$ square, while Cs layers intercalate between neighboring $V_2Se_2O$ layers. Within the spin-up (down) sublattice, the V-O bonds align along the $x$ ($y$) direction (Fig. 1b). Connecting the opposite spin sublattices requires a non-relativistic spin group operation, $[C_2||C_{4z}]$, which combines a 180° rotation in spin-space ($C_2$) with a 90° real-space rotation about the z-axis ($C_{4z}$). This operation reduces the symmetry of each spin sublattice from $C_4$ to $C_2$. Previous Angle-resolved photoemission spectroscopy (ARPES) and nuclear magnetic resonance (NMR) studies (*28*) report two distinct magnitudes of magnetic moment for both spin-up and spin-down, forming a √2

× √2 spin density wave (SDW) in the $V_2O$ plane (Fig. 1b), supported by a density-wave-like transition near 100 K observed in transport measurements (Extended Data Fig. 2)[40]. The spin-defined environment from the four surrounding V atoms dictates the electronic state of each surface Se atom, thus driving the formation of a √2 × √2 charge density wave (CDW, dashed box in Fig. 1b). Our STM measurements directly visualize this SDW-induced CDW on the cleaved surface (Fig. 1d, e). The inset of Fig. 1d shows both Bragg peaks and √2 × √2 CDW peaks in the fast Fourier transformation (FFT) result. The measured lattice constant ($a$ = 4.07 ± 0.05Å) agrees with the reported values for $CsV_2Se_2O$ single crystals[35]. Uniform d$I$/d$V$ spectra acquired across the clean Se surface show a consistent SDW gap of ~ 70 meV (upper panel of Fig. 1f). Some previous studies also explored the origin of the density wave in the $AV_2Se_2O$ system[40,41], our results are in excellent agreement with the previous ARPES and NMR studies[28] and provide atomic-scale evidence for the SDW. Our DFT calculations confirm the spin-splitting and the altermagnetic nature of $CsV_2Se_2O$, regardless of the presence of the √2 × √2 SDW (Extended Data Fig. 3).

In a large-scale STM topographic image (Fig. 1e), strip-like features (yellow arrows) and two types of defects are observed on the Se surface. These strip-like features correspond to spin-defect lines along the (110) direction, whose detailed structure is discussed later. Both Defect 1 and Defect 2 exhibit a cross-shaped morphology (see Extended Data Fig. 4), with Defect 1 located on the spin-defect lines and Defect 2 off the lines. d$I$/d$V$ spectra acquired at Defect 1 and Defect 2 reveal in-gap states at distinct energies (~ 20 meV for Defect 1, ~ 40 meV for Defect 2, lower panel of Fig. 1f). As discussed below, these are both spin defects that enable direct imaging of the rotational-symmetry-breaking states of altermagnetism.

**Rotational-symmetry-breaking states**

Real-space rotational-symmetry breaking serves as conclusive proof of altermagnetism. The first rotational-symmetry-breaking state is observed around Defect 1. Defect 1 induces a quasi-one-dimensional (quasi-1D) static charge order that propagates into neighboring areas along either the $x$ or $y$ direction (Fig. 2a, b). This charge order, with a period of $2a_0$ (twice the Se square lattice constant), aligns with both the $C_2$ symmetry and the V-O bond direction of the host spin sublattice. In Fig. 2a, spin-defect lines manifest as dark strips, while in the lower-energy d$I$/d$V$ map (Fig. 2b), they appear as bright strips. This contrast reversal confirms the electronic, instead of structural, origin of the stripes. The static charge orders exhibit identical orientations on the same spin-defect line but adopt orthogonal directions on adjacent lines (double-headed arrows, Fig. 2b). These findings are reproducible across different samples and at various locations on the same sample (Extended Data Fig. 5).

d$I$/d$V$ maps at two characteristic energies reveal both defect-induced static order and QPI patterns (Fig. 2b, c). Their corresponding FFT results reveal three wavevectors:

the static charge order wavevector $q_0$ (induced by Defect 1, Fig. 2d), and the QPI wavevectors $q_1$ and $q_2$, presented in Fig. 2d and 2e, respectively. $q_1$ and $q_2$ exhibit distinct arc-like shapes, convex for $q_1$, and concave for $q_2$, along with different energy dependencies. Inverse fast Fourier transformation (IFFT) results demonstrate that the real-space signals of $q_0$ and $q_1$ are strictly colocalized around Defect 1 on the spin-defect lines, with $q_0$ appearing straight and $q_1$ rounded (Fig. 2f, g). In contrast, the $q_2$ is confined around Defect 2 off the spin-defect lines (Fig. 2h). This spatial distribution indicates that line-bound Defect 1 simultaneously induces the quasi-1D static order ($q_0$) and acts as the scattering center for $q_1$, while Defect 2 off the lines is the exclusive scattering center for $q_2$. d$I$/d$V$ maps and corresponding FFT results at more energies are shown in Extended Data Fig. 6 and 7.

The unidirectional nature of all observed static order and QPI patterns (Fig. 2f-h) directly demonstrates rotational-symmetry breaking in real space. This is consistent with their origin in the material's *d*-wave altermagnetism, which we analyze in the following section.

Another symmetry-breaking state emerges as a charging ring induced by Defect 2 off the spin lines. We first characterize the structure of Defect 2. STM topography (Fig. 3a) reveals two typical forms of Defect 2: one with a smaller and one with a larger apparent size. The schematic in Fig. 3b depicts the cross-shaped, smaller variant, featuring a central Se vacancy on the surface and associated spin defects on two nearest-neighbor V-atoms (blue dashed circles). Given that each $V_2O$ layer is sandwiched between two Se layers (Fig. 1a), equivalent spin defects can be induced by a Se vacancy in the bottom layer. We attribute the larger, checkerboard-like Defect 2 in Fig. 3a to such a bottom-layer vacancy (Extended Data Fig. 8), analogous to the case of Fe vacancy in FeSe[42]. The defect site itself does not break any rotational symmetry. Consistently, both the morphology and d$I$/d$V$ maps show well-preserved $C_4$ symmetry of the defects (see also Extended Data Fig. 4).

A series of ellipse-shaped charging rings is clearly resolved in d$I$/d$V$ maps at the characteristic energy of Defect 2 (Fig. 3c). These $C_2$-symmetric ellipses have their long axes aligned along either the $x$ or $y$ direction, matching the symmetry of the spin-up or spin-down sublattices (Fig. 3b). To highlight their spatial relationship with the defects, the rings are superimposed on a current map where only defects are visible (Fig. 3d). All rings are localized around Defect 2 and confined between neighboring spin-defect lines, never crossing them. This indicates that the rings are induced by the spin defects of Defect 2 and inherit the symmetry of the host spin sublattice. The rings are displaced from the center of Defect 2, likely due to the inequivalence of the two V-atom spins associated with the Se vacancy (Fig. 3b). The coexistence of rings with long axes along both $x$ and $y$ in Fig. 3c confirms that the $C_2$ symmetry is an intrinsic material property, ruling out artifacts from the STM tip shape. Charging rings in semiconductors have been widely studied[43-45] and typically exhibit an isotropic annular shape. The anisotropy of the elliptical charging rings observed here is a unique characteristic of altermagnetic systems. Between the neighboring spin-defect lines, rings with orthogonal orientations can both be observed (Fig. 3d). This can be understood from the spin sublattice configuration: each Se vacancy is surrounded by two spin-up and two spin-down V atoms, with the two V-atom pairs being mutually perpendicular. Consequently, defects (and thus their associated charging rings) linked to both spin sublattices can form. In some defect-rich areas, the orthogonal charging rings may even intersect (Extended Data Fig. 9). This further confirms the correspondence between the spin sublattice and the charging ring directions. The coexistence of orthogonal charging rings provides a firm basis for discussing the origin of the rotational-symmetry-breaking states in the altermagnetism framework; as we show below, it offers strong real-space evidence for altermagnetism.

In addition to the cross-shaped defect, dimer-shaped defects are also observed. Extended Data Fig. 10 shows that all the previously mentioned phenomena, including the quasi-1D static charge order, elliptical charging ring and QPI pattern, can be equally induced by dimer-shaped defects. It follows that these intriguing phenomena are intrinsic to altermagnetism and are largely independent of the defect shape.

**Origin of ellipse charging ring**

Figure 3e, f present close-up d$I$/d$V$ maps of the charging ring around Defect 2. Spatially resolved spectroscopy along the ring's long axis reveals that the defect-state peak at ~40 meV (also in Fig. 1f) shifts to lower energy with increasing lateral distance from the defect (Fig. 3g, h). This shift is consistent with the tip-induced band bending[39,43-45]: a downward band bending moves the defect state, initially slightly above $E_F$ within the SDW gap, below $E_F$ (Fig. 3i). The magnitude of band bending increases with smaller tip bias ($V$) and smaller tip-defect distance. The tunneling process between tip and localized defect state contributes to the observed charging ring. Consequently, in a d$I$/d$V$ map at a fixed bias, the charging ring appears at the lateral position where the defect state $E_D$ is bent across $E_F$ (Fig. 3i); a lower bias produces a larger ring (Extended Data Fig. 11). Conversely, at a greater lateral distance, a lower bias is required to achieve the same band bending, explaining the observed energy shift in the spectra (Fig. 3g, h). Notably, the peak shifts more rapidly when moving away from Defect 2 along the short axis of the ring (Extended Data Fig. 12), indicating that the band bending decays faster along one principal direction ($x$ or $y$) than the other. As shown in Fig. 1c, the quasi-1D spin-up $\alpha$ band disperse primarily along $k_x$ while being almost dispersionless along $k_y$. Therefore, band bending of the corresponding defect state decays more slowly along the $x$ direction, producing an elliptical charging ring with the long axis aligned along the $x$ direction.

This directional anisotropy directly reflects the rotational symmetry breaking of the charging rings, which fundamentally stems from the $C_2$ symmetry of the spin sublattice and the corresponding anisotropic band structure.

**Origin of unidirectional static order**

We now examine the unidirectional static charge order that emerges along either the *x* or *y* direction around Defect 1 on the spin-defect lines. A zoom-in d*I*/d*V* map reveals its characteristic W-shaped profile (Fig. 4a). The proposed atomic structure (Fig. 4b) indicates that, for charge order along *x*, a spin-down V-O chain lies at the center of each blue W-shaped shaded area; for order along *y*, a spin-up V-O chain centers the light red shaded W-shaped area. The W-shaped pattern results from the superposition of the V-O chains (period $2a_0$) and the $\sqrt{2}\times\sqrt{2}$ CDW on the Se layer. Like the charging rings, this order originates from spin defects, and its quasi-1D character likewise stems from the $C_2$ symmetry of the spin sublattices.

Critically, all quasi-1D static charge orders on the same spin-defect line share the same orientation (Fig. 2f). We propose that a given spin-defect line hosts only one type of spin defect (spin-up *or* spin-down). As depicted in Fig. 4b, a spin-defect line is formed by two columns of spin-down V atoms. Consequently, only spin-down defects can be induced by a surface Se vacancy on that line. Such a spin-down defect on the line highlights the spin-down V-O chain in the neighboring area, leading to static charge order exclusively along *x* (blue shaded area). This is consistent with the calculation result that the electronic states around a defect have the same $C_2$ symmetry of the spin sublattice in which the defect is located[46]. Importantly, defects on adjacent spin-defect lines always induce orthogonal unidirectional charge orders (Fig. 2f and Extended Data Fig. 5), indicating that these adjacent spin-defect lines are composed of opposite spins. This points to a long-range antiferromagnetic interaction between the spin-defect lines. Moreover, the spacing between the spin-defect lines is nearly uniform, suggesting the possible emergence of a new long-range SDW order that warrants further investigation.

STS spectra along the charge order reveal intriguing electronic states near 20 meV (Fig. 4d, e). The defect state of Defect 1 splits into two branches away from the defect site (Fig. 4c-e). The unsplit center of this state is also offset from the Se-vacancy site (red cross and dashed line in Fig. 4c, e), a displacement attributed to the inequivalence of the two V-atom spins associated with the vacancy. This spatial distribution is not a tip-induced artifact, as confirmed by spectra taken at fixed locations with varying setpoints (Extended Data Fig. 13).

**Origin of observed QPIs**

The convex $q_1$ and concave $q_2$ wavevectors (Fig. 2d, e and Fig. 5a, f) characterize the distinct QPI patterns. To understand their origin, we performed self-correlation simulations based on the spin-resolved Fermi surface (FS) from ARPES[28] (Fig. 5b). All the simulation results are in the same scale with the QPI patterns in Fig. 5a, f. The simulation considering only spin-conserved scattering (Fig. 5c) reproduces the observed wavevector $q_2$ perfectly in both shape and length. In contrast, the simulation that includes spin-flip processes (Fig. 5d) yields additional features, as highlighted by the gray circles. Crucially, these extra features are absent in our experimental QPI data (Fig. 5a). This agreement demonstrates conclusively that $q_2$ arises from spin-conserved scattering between FS contours of the same spin orientation (red arrow in Fig. 5b). While magnetic impurities allow spin-flip processes, the initial and final spin states remain orthogonal in *d*-wave altermagnet and contribute no interference, analogous to the case of topological insulators[47-50]. $q_2$ is observable in QPI because its energy lies near the SDW gap edge. The measured dispersion of $q_2$ (from 35-65 meV) exhibits electron-like behavior (Fig. 5e), consistent with both band structure calculations and ARPES data[28] (Extended Data Fig. 14).

The QPI wavevector $q_1$ (Fig. 5f), coexisting with the static charge order $q_0$, is distinct from $q_2$ in both real-space location and energy, appearing around Defect 1 within the 16-36 meV range (Fig. 2f, g). To understand $q_1$, the FS folding induced by the $q_0$ charge order (period $2a_0$, Fig. 5g) must be considered. However, self-correlation simulations (Fig. 5g inset) based on this folded FS fail to reproduce the observed $q_1$.

A consistent picture emerges by considering the energy hierarchy of the gap-opening: while the $\sqrt{2}\times\sqrt{2}$ SDW opens a uniform gap across the FS[28], the static charge order ($q_0$) introduces a secondary gap only at higher energy (Fig. 5h). Therefore, within the SDW gap at the energies relevant to $q_1$, the electronic structure is effectively governed by the bands folded solely by the static charge order (bands and FS contours denoted by dashed lines in Fig. 5h, i). Simulations based on this picture (Fig. 5j) successfully reproduce $q_1$. The extracted dispersions (Fig. 5k) confirm the non-dispersing nature of $q_0$. As for $q_1$, its appearance within a narrow energy range precludes a definitive dispersion analysis.

While the spin-conserved nature of the quasiparticle scattering, consistent with a prior report[7], is confirmed by our analysis, the key advance of our work lies in the direct real-space observation of rotational symmetry breaking. Although the Fourier transformations of $q_1$ and $q_2$, which spatially average signals from multiple defects, exhibit an apparent $C_4$ symmetry (Fig. 2d, e and Fig. 5a, f), the QPI maps around individual defects reveal clearly unidirectional interference patterns (Fig. 2g, h). This defect-resolved unidirectionality, together with the other symmetry-breaking electronic states—the static charge order and the anisotropic charging rings—provides unambiguous real-space evidence for altermagnetism, moving beyond the established momentum-space signatures.

**Outlook**

Our atomic-scale visualization of altermagnetism not only confirms the long-predicted rotational symmetry breaking, but further reveals how this fundamental symmetry lowering manifests in distinct unidirectional electronic states. These real-space patterns encode the underlying alternating spin texture and its coupling to the lattice, offering a new window into the microscopic interplay between spin, charge, and crystal symmetry in altermagnets. The observed long-range antiferromagnetic coupling between adjacent spin-defect lines hints at an emergent spin order, opening further questions about its origin and relation to the altermagnetic ground state.

$CsV_2Se_2O$ thus emerges as a clean prototype for *d*-wave altermagnetism. Intriguingly, symmetry-breaking electronic orders resembling those observed here are also found in cuprate superconductors, where they are linked to the pseudogap [51,52] and unconventional superconductivity[53]. In $CsV_2Se_2O$, however, the band structure with *d*-wave spin-splitting originates directly from crystal symmetry, providing a clean, correlation-light platform to disentangle symmetry-driven phenomena from strong correlation effects. The intimate coupling between altermagnetism and charge degrees of freedom suggests that related materials may host emergent phases such as spin-triplet superconductivity. A promising route is to interface such *d*-wave altermagnets with superconductors[54,55], where the spin-split Fermi surfaces could stabilize equal-spin pairing and thus generate spin-triplet superconductivity, offering a materials-realizable pathway to this long-sought state.

*Note added* – While submitting the manuscript, we became aware of a related work, which supports the altermagnetism in $KV_2Se_2O$ by quasiparticle interference observed in spin-polarized STM measurements[56].

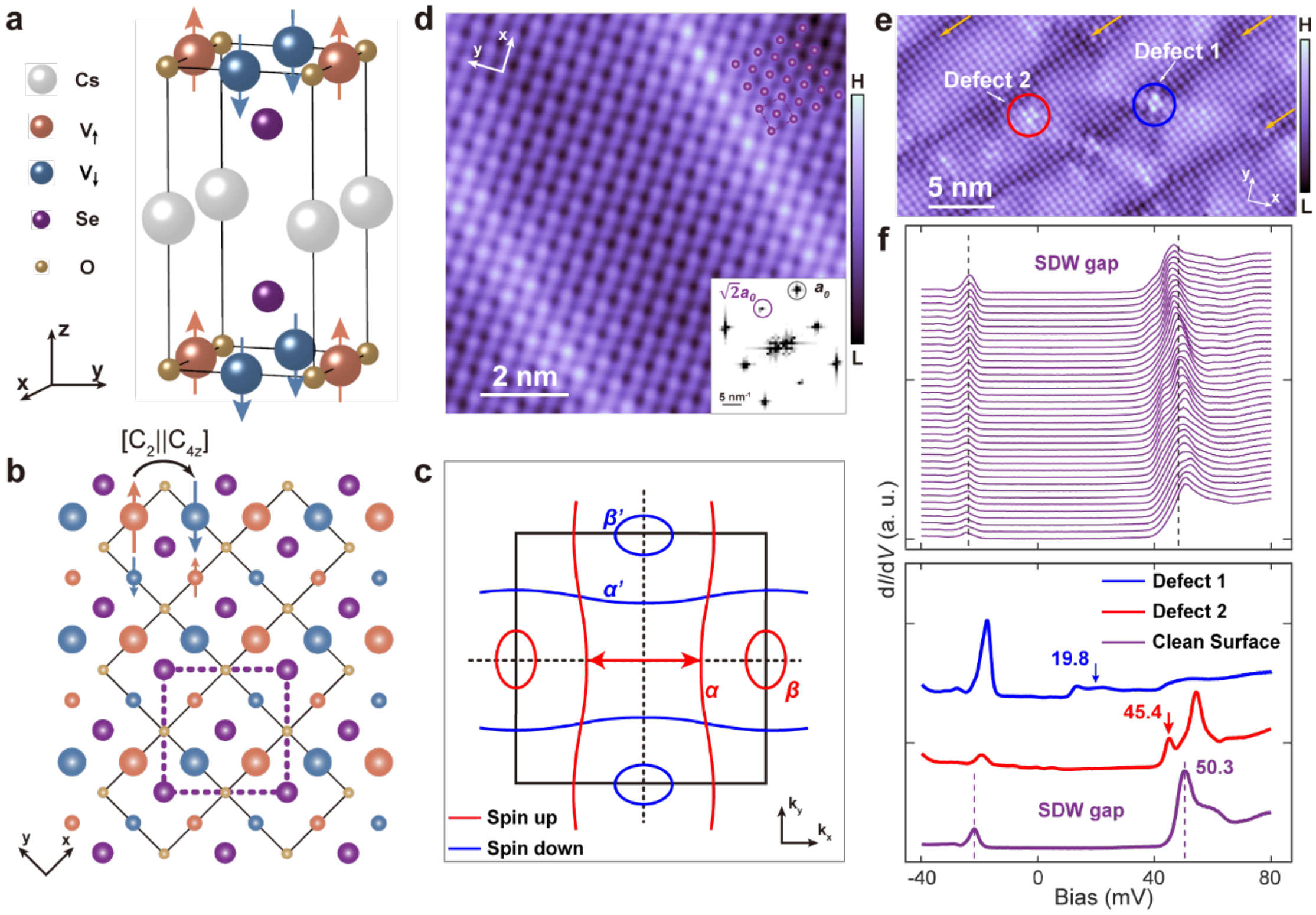

**Fig. 1 | Spin density wave gap and induced charge order in the altermagnet $CsV_2Se_2O$. a,** Crystal and magnetic structure of $CsV_2Se_2O$. Red and blue arrows indicate the up and down spins of the V atoms. **b,** Top view of the Se plane and $V_2O$ plane. The size of the blue and red spheres indicates the magnitudes of the magnetic moments of the V atoms. Purple dashed square denotes one unit cell of the SDW-induced $\sqrt{2} \times \sqrt{2}$ CDW on the Se surface. The opposite spin lattice is connected by the $[C_2||C_{4z}]$ operation. **c,** Schematic Fermi surface with red (blue) lines indicating the constant energy contour of spin-up (down) electrons. Black lines indicate the first Brillouin zone. Red double-headed arrow indicates the spin-conserved scattering process. **d, e,** Atomically resolved STM topographic image of a cleaved Se surface (**d**: 9 nm × 9 nm, set point $V_s$ = 30 mV, $I_t$ = 200 pA; **e**: 30 nm × 15 nm, 60 mV, 800 pA). Purple spheres in **d** denote the Se square lattice. Inset of **d**: the fast Fourier transformation result of **d**, gray and purple circles denote the Bragg peak of the Se square lattice and the $\sqrt{2} \times \sqrt{2}$ CDW, respectively. Yellow arrows in **e** indicate spin-defect lines. Blue and red circles in **e** indicate a defect on the spin-defect line (Defect 1) and a defect off the line (Defect 2), respectively. **f,** Top: A series of $dI/dV$ spectra (set point $V_s$ = 80 mV, $I_t$ = 800 pA) taken on the clean surface. Black dashed lines indicate a uniform SDW gap. Bottom: $dI/dV$ spectra (80 mV, 800 pA) taken on clean surface, Defect 1 and Defect 2, respectively. Purple dashed lines indicate the SDW gap on the clean surface, blue and red arrows denote defect-induced electronic states.

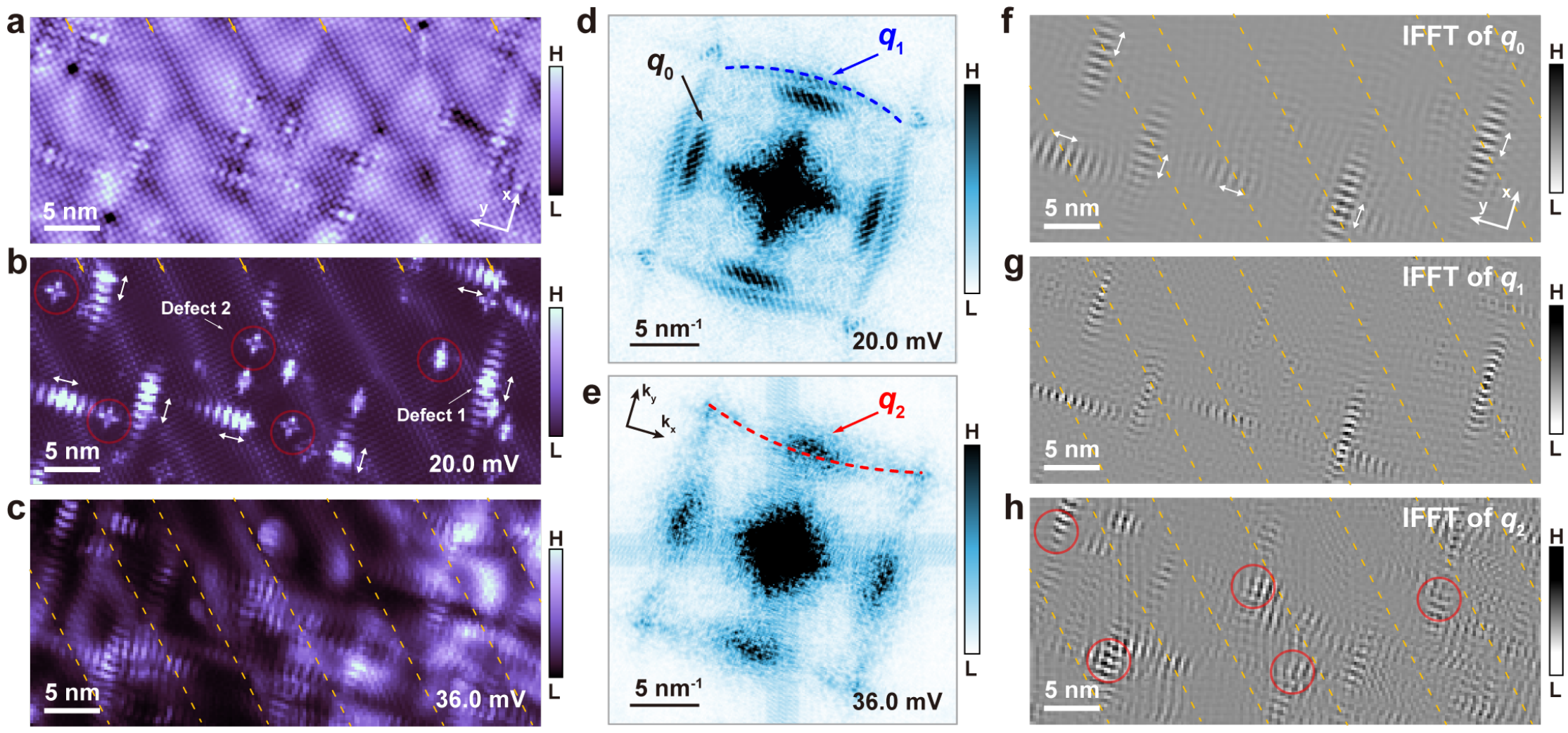

**Fig. 2 | Quasi-1D static charge order and quasiparticle interference pattern in $CsV_2Se_2O$. a,** STM topographic image of a Se surface (45 nm × 20 nm, set point $V_s$ = 50 mV, $I_t$ = 500 pA). **b, c,** d$I$/d$V$ map of the Se surface (45 nm × 20 nm, set point $V_s$ = 80 mV, $I_t$ = 1.2 nA) at the energy of 20.0 mV (**b**) and 36.0 mV (**c**). Red circles in **b** indicate Defect 2, white double-headed arrows indicate Defect 1 and the quasi-1D static charge order around it. Yellow arrows and dashed lines in **a**, **b**, **c** denote the spin-defect lines. Static charge orders are aligned identically on the same spin-defect line but orthogonally on neighboring lines. **d, e,** The symmetrized Fast Fourier transformation result of **b**(**d**) and **c**(**e**). Black arrow indicates the static charge order wavevector $q_0$, blue (red) arrow and dashed line indicate the quasiparticle interference wavevector $q_1$ ($q_2$). **f, g, h,** Inverse fast Fourier transformation result of static charge order wavevector $q_0$ (**f**), QPI wavevector $q_1$ (**g**) and $q_2$ (**h**). Yellow dashed lines, white double-headed arrows and red circles retain their meanings as defined in **a**-**c**. The scattering centers of $q_1$ and $q_2$ are Defect 1 (on the spin-defect line) and Defect 2 (off the line), respectively. The $q_2$-related scatterings are marked red circles.

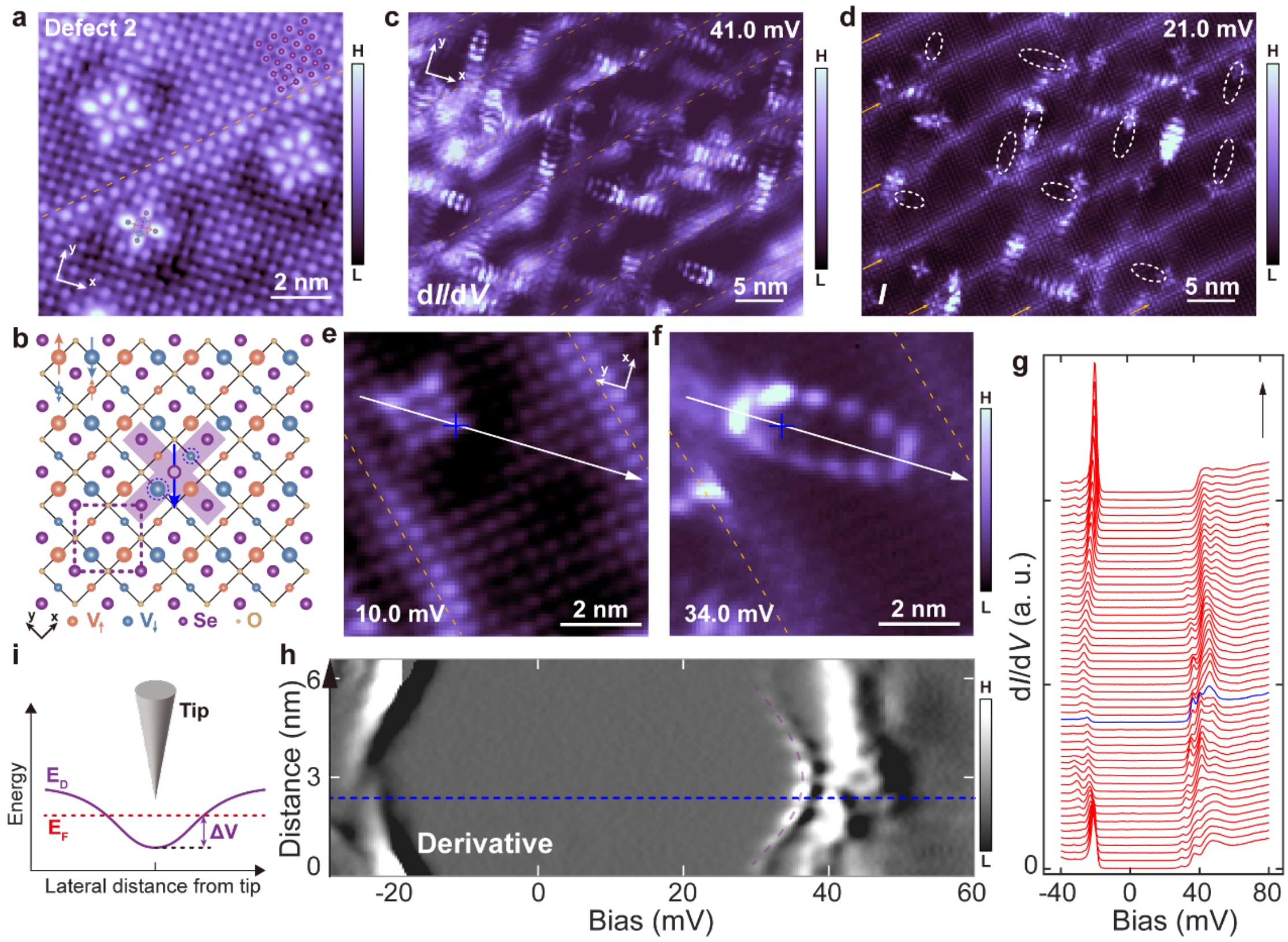

**Fig. 3 | Charging ring with broken rotational symmetry in altermagnetic state. a,** STM topographic image of Defect 2 (10 nm × 10 nm, $V_s$ = 10 mV, $I_t$ = 200 pA). Purple spheres denote the Se square lattice. **b,** Atomic structure of Defect 2. The cross-shaped purple shaded area corresponds to the composite defect site (atoms assembled in **a**), which is defined by a central Se vacancy (purple circle) and a spin defect on its nearest-neighbor V atoms (two blue dashed circles). A purple dashed square denotes a unit cell of the √2×√2 CDW on Se surface. Two additional defects with larger apparent sizes are of the same type as Defect 2 but reside in deeper atomic layers (Extended Data Fig. 6). **c,** d$I$/d$V$ map (40 nm × 30 nm, 60 mV, 1.2 nA) at the energy of 41.0 mV. Rotational-symmetry-breaking charging rings are clearly visualized. **d,** Current map (40 nm × 30 nm, 60 mV, 1.2 nA) at an identical area with **c** at the energy of 21.0 mV. White dashed circles indicate the charging rings in **c**. **e, f,** d$I$/d$V$ maps of the charging ring around Defect 2 (7.5 nm × 7.5 nm, 60 mV, 1.2 nA) at the energy of 10.0 mV (**e**) and 32.0 mV (**f**). Yellow dashed lines and arrows in **a**-**f** indicate the spin-defect lines. **g,** A series of STS spectra taken along the white arrow in **e** and **f** (80 mV, 800 pA). **h,** False color image of the derivative of the d$I$/d$V$ spectra in **g**. Blue line and dashed line in **g** and **h** indicate the spectrum taken at the blue cross position in **e** and **f**. **i,** Schematic of tip-induced band bending. Purple (dashed) lines in **i** and **h** indicate the bended defect state $E_D$.

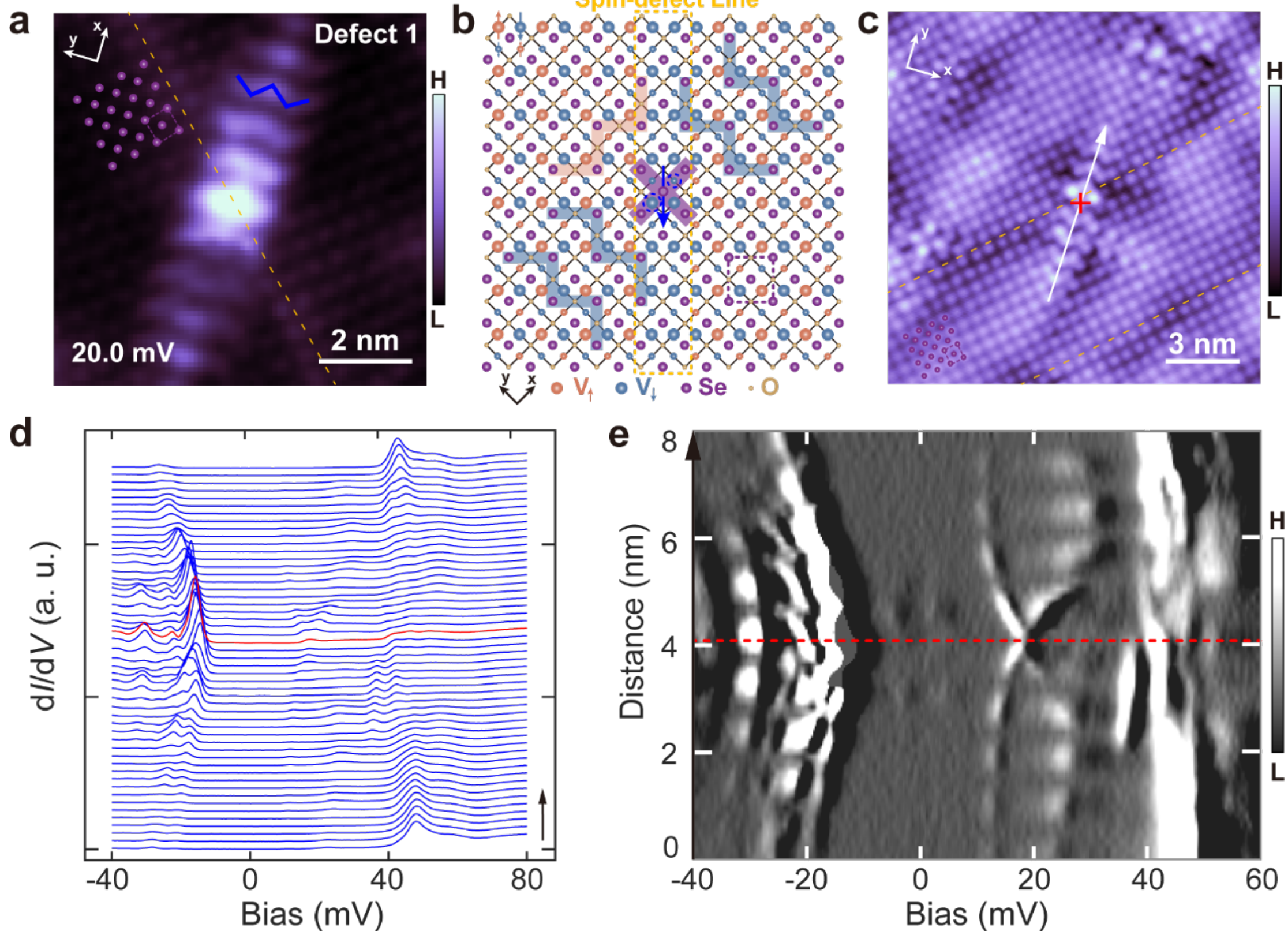

**Fig. 4 | Unidirectional electronic states induced by defect on the altermagnetic spin-defect line. a,** d$I$/d$V$ map (10 nm × 10 nm, set point $V_s$ = 60 mV, $I_t$ = 1.2 nA) of quasi-1D static charge order induced by a defect on the spin-defect line (Defect 1) at the energy of 20.0 mV. Yellow dashed line indicates the spin-defect line. W-shaped blue line denotes the static charge order. Purple spheres denote the Se square lattice. **b,** Atomic structure of the altermagnetic spin-defect line. The size of the red (blue) spheres indicates the magnitude of the magnetic moments of the spin-up (down) of V atoms. Yellow dashed rectangle indicates the altermagnetic spin-defect line. W-shaped blue shaded areas indicate the surface highlighted Se atoms corresponding to the quasi-1D static charge order in **a**. W-shaped light red shaded area indicates the possible static charge order in the orthogonal direction. Purple dashed lines indicate a unit cell of the $\sqrt{2} \times \sqrt{2}$ SDW-induced CDW on the Se surface. **c,** STM topographic image of quasi-1D static charge order (10 nm × 10 nm, set point $V_s$ = 60 mV, $I_t$ = 1.2 nA). Yellow dashed lines and purple spheres retain their meanings in **a**. **d,** A series of d$I$/d$V$ spectra taken along the white arrow in **c** (set point $V_s$ = 80 mV, $I_t$ = 800 pA). **e,** False color image of the derivative of the d$I$/d$V$ spectra in **d**. Red (dashed) line in **d** and **e** indicates the spectrum taken at the point denoted by the red cross in **c**.

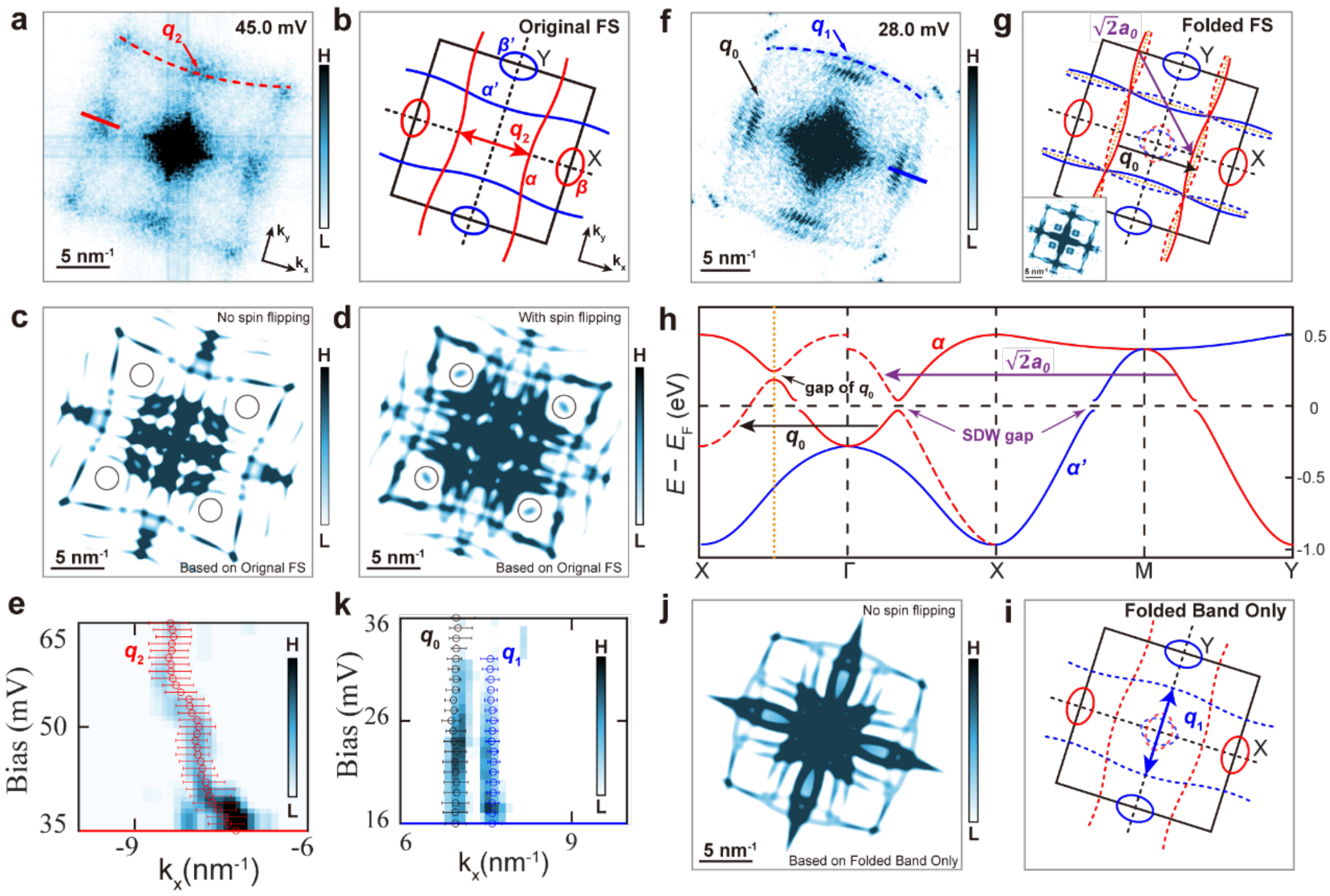

**Fig. 5 | Origin of observed QPI in spin-conserved scattering process. a,** Symmetrized fast Fourier transformation of d$I$/d$V$ map (45 nm × 20 nm, $V_s$ = 80 mV, $I_t$ = 1.2 nA) at 45.0 meV. Red dashed line and arrow indicate the QPI wavevector $q_2$. **b,** Schematic Fermi surface with red (blue) lines indicating the constant energy contour of spin-up (down) electrons. Black lines indicate the first Brillouin zone. Red double-headed arrow indicates the spin-conserved scattering process for $q_2$. **c, d,** Simulated QPI wavevectors based on the Fermi surface in **b** with (**d**) and without (**c**) spin flipping process. **e,** Dispersion of $q_2$ extracted from the red cut in **a**. Red circles with error bars indicate the measured wavevectors at each energy. **f,** Symmetrized fast Fourier transformation of d$I$/d$V$ map (42 nm × 42 nm, 60 mV, 1.2 nA) at 28.0 meV. Black and blue arrows indicate the static order wavevector $q_0$ and QPI wavevector $q_1$. **g,** Fermi surface folded by $q_0$ with red (blue) dashed lines indicating the constant energy contour of folded band of spin-up (down) electrons. Orange dashed line indicates the folding boundary of $q_0$. Inset: Simulated QPI wavevectors based on **g**. **h,** Schematic of folded bands. Red and blue (dashed) lines indicate spin-up and spin-down (folded) energy bands. Black and purple arrows in **g** and **h** indicate the charge order wavevector $q_0$ and the SDW wavevector. Only one folded branch (spin-up) from $q_0$ and SDW is plotted for clarity. **i,** Fermi surface with only folded bands. Blue double-headed arrow indicates spin-conserved scattering process corresponding to $q_1$. **j,** Simulated QPI wavevectors based on **i** without the spin flipping process. **k,** Dispersion of $q_0$ and $q_1$ extracted from the blue cut in **f**. Black (blue) circles with error bars indicate the measured wavevectors $q_0$ ($q_1$) at each energy.

## Methods

### Sample preparation

Single crystals of $CsV_2Se_2O$ were grown using a flux method with CsSe as the flux. The starting materials of Cs, Se, V, and $V_2O_5$, mixed in a molar ratio of $CsV_2Se_2O$: CsSe = 1:5, were loaded into an alumina crucible. The crucible was sealed in an evacuated quartz tube and pre-reacted at 40 °C for 1 hour. Subsequently, the quartz tube was heated slowly to 1000 °C, at which held for 10 hours to ensure homogenization, and then slowly cooled to 650 °C at a rate of 2 °C/h. Finally, the $CsV_2Se_2O$ crystals were separated from the excess flux by centrifugation.

### STM Measurements

STM measurements were carried out in a low-temperature STM system (Unisoku 1200). A polycrystalline PtIr STM tip was calibrated on Ag island. STS data were taken by standard lock-in method. The feedback loop is disrupted during data acquisition and the frequency of oscillation signal is 811.0 Hz. $CsV_2Se_2O$ single crystal were cleaved at 77 K under ultra-high vacuum (with the pressure $< 2\times10^{-10}$ Torr) and then transferred to STM head at 4 K.

### QPI Simulation

The spin-resolved Fermi surface (Fig. 5b) was constructed in accordance with prior ARPES measurements and theoretical calculations[28]. For the simulation, we considered only constant-energy contours within the first Brillouin zone and assumed uniform intensity across the Fermi surface. Simulated QPI patterns were generated by calculating the self-correlation of the reproduced Fermi surface. In simulations without spin-flip processes (Fig. 5c, j), only scattering between contours of the same spin orientation was included. For simulations that include spin-flip scattering, self-correlation was performed over the full Fermi surface without distinguishing spin textures (Fig. 5d, inset of 5g).

### Electrical Transport and Magnetic Measurements

Electrical transport was measured using a physical property measurement system (PPMS, Quantum Design) from 300 K down to 1.8 K. A rectangular crystal, cut with a wire saw, was contacted in a standard four-probe configuration using 30 μm gold wires bonded with room-temperature-cured silver paste. Magnetization as a function of temperature and magnetic field was acquired with a magnetic property measurement system (MPMS-VSM, Quantum Design) between 300 K and 2 K in fields up to ±7 T.

### DFT calculation

All calculations were performed within the framework of density functional theory (DFT) as implemented in the Vienna *ab initio* simulation package (VASP). We employed the projector augmented wave (PAW) method with a plane-wave basis set. The exchange-correlation functional was treated using the generalized gradient approximation (GGA) in the Perdew–Burke–Ernzerhof (PBE) parameterization. An energy cutoff of 600 eV was set for the plane-wave basis. Brillouin zone integration was carried out using a $9\times9\times5$ $\Gamma$-centered k-point mesh generated by the Monkhorst-Pack scheme. During geometry optimization, lattice constants of the $CsV_2Se_2O$ unit cell were fixed to the experimental values ($a = b = 4.01$ Å, $c = 7.871$ Å), while the atomic positions were relaxed until the magnitude of each Hellmann–Feynman force component was less than 0.001 eV/Å. The self-consistent field calculations were considered converged when the total energy change between successive iterations fell below $10^{-7}$ eV. To account for the strong electron correlation in the V-3*d* orbitals, the GGA+U (Hubbard U) method was employed. The effective on-site Coulomb parameter U was assigned based on the crystallographic environment of V atoms: the larger V sites were assigned U = 1 eV, while the smaller sites were treated with U = 0 eV.

**Data availability**

All study data are included in the article and/or Extended Data.

**Acknowledgements**

D.F. and L.Y. contributed equally to this work. We thank C. Wu, Z. Lu, X. Wan, Z. Zhang, H. Weng, H. Lu, G. Jia, J. Yan, Y. Peng, Z. Hu and W. Dong for helpful discussions. The experimental work was supported by the National Natural Science Foundation of China (NNSFC, Grant No. 92365201), the National Key R&D Program of China (Grant No. 2022YFA1403103), NNSFC (Grants No. 12321004, No. 52388201, and No. 11427903), and Quantum Science and Technology-National Science and Technology Major Project (2021ZD0302402).

**Author Contributions**

W.L. and Q.-K.X. conceived the research project. D.F. and K.X. performed the STM experiments with the assistance of Y.W.. L.Y., Z.W. and Y.Y. grew the samples. W.J. and Y.S. performed DFT calculations. W.L., D.F. and K.X. analyzed the data. W.L., D.F. wrote the manuscript with input from all other authors.